 \documentclass[12pt,preprint]{aastex}

\usepackage{graphicx}
\usepackage{amsmath}

\def \apj {ApJ}

\def \solphys {Solar Phys.}

\def \aap {A\&A}

\newcommand{\citeN}[1]{\citeauthor{#1} (\citeyear{#1})}
\newcommand{\citeNP}[1]{\citeauthor{#1} \citeyear{#1}}

\shortauthors{Socas-Navarro}
\shorttitle{Stokes Inversion with Neural Networks}

\begin{document}

\title{Strategies for Spectral Profile Inversion using Artificial Neural
  Networks} 

\author{H. Socas-Navarro}
   	\affil{High Altitude Observatory, NCAR\thanks{The National Center
	for Atmospheric Research (NCAR) is sponsored by the National Science
	Foundation.}, 3450 Mitchell Lane, Boulder, CO 80307-3000, USA}
	\email{navarro@ucar.edu}

\date{}%

\begin{abstract}
This paper explores three different strategies for the inversion of spectral
lines (and their Stokes profiles) using artificial neural networks. It is
shown that a straightforward approach in which the network is trained with
synthetic spectra from a simplified model leads to considerable errors in the
inversion of real observations. This problem can be overcome in at least two
different ways that are studied here in detail. The first method makes use of
an additional pre-processing auto-associative neural network to project the
observed profile into the theoretical model subspace. The second method
considers a suitable regularization of the neural network used for the
inversion. These new techniques are shown to be robust and reliable when
applied to the inversion of both synthetic and observed data, with errors
typically below $\sim$100~G. 
\end{abstract}


\section{Introduction}
\label{sec:intro}

The analysis of the spectral properties of the light intensity and its
polarization state is the basis of modern solar physics. For over two
decades, least-squares profile fitting has been the method of
choice (see \citeNP{SN01a}; \citeNP{dTI03} and references therein). Many
different inversion codes, based on a variety of physical models, have been
developed and used extensively for the determination of magnetic and
thermodynamic conditions in the atmosphere.

While the use of least-squares fitting has important benefits, there is an
increasing demand for alternative procedures that are faster and more
robust. By robust I mean capable of operating reliably without human
intervention on a routine basis. This demand is driven by the development of
a new generation of spectro-polarimeters, which will deliver enormous data
flows (SOLIS, \citeNP{K98}; Solar-B, \citeNP{LES01}; Sunrise,
\citeNP{SBK+03}; DLSP, \citeNP{SEL+03}). 

The attention of solar physicist has turned in recent years towards a new
breed of diagnostic techniques based on pattern recognition and machine
learning. A considerable number of papers has been devoted to the
investigation of these techniques (\citeNP{RLAT+00}; \citeNP{SNLAL01};
\citeNP{CS01}; \citeNP{SLA02}; \citeNP{SN03}; \citeNP{dTILA03};
\citeNP{SN04d}). Most of those works deal with the Principal Component
Analysis, which is a series expansion of the spectra. Another line of
research explores the use artificial neural networks (ANNs), which show
considerable promise for the inversion of spectral observations
(\citeNP{CS01}; \citeNP{SN03}). 

The work presented here demonstrates the applicability of ANNs to actual
observed data in typical working conditions. The radiative transfer
computations that are needed to synthesize training profiles have been
carried out using the Milne-Eddington approximation. This is helpful to
simplify the problem and to allow for the synthesis of thousands of profiles
in a reasonable computing time. All the calculations done for this work, and
the CPU times quoted below, were obtained using a Pentium~IV processor
running at 1.2~GHz. The ANN inversions require very little computational
resolurces both in terms of processor speed and memory storage. The training
algorithm, on the other hand, can be very demanding. 

\section{The neural network}
\label{sec:ANN}

An ANN is a structure of interconnected
neurons, where each neuron is a memory cell with the capability to store a
real number. The number stored in a neuron can be modified according to the
contents of its neighbors and the ``synaptic weights'' that connect
each pair of neurons. A perturbation in the contents of a neuron propagates
through the network following the structure of connections and 
synaptic weights. For a sake of tractability, it is customary to consider
{\it forward-propagating} networks. These ANNs have a well defined signal
propagation direction, starting at a set of input neurons and ending at
the outputs. No feed-back loops are allowed in forward-propagating networks.

\begin{figure*}
\plotone{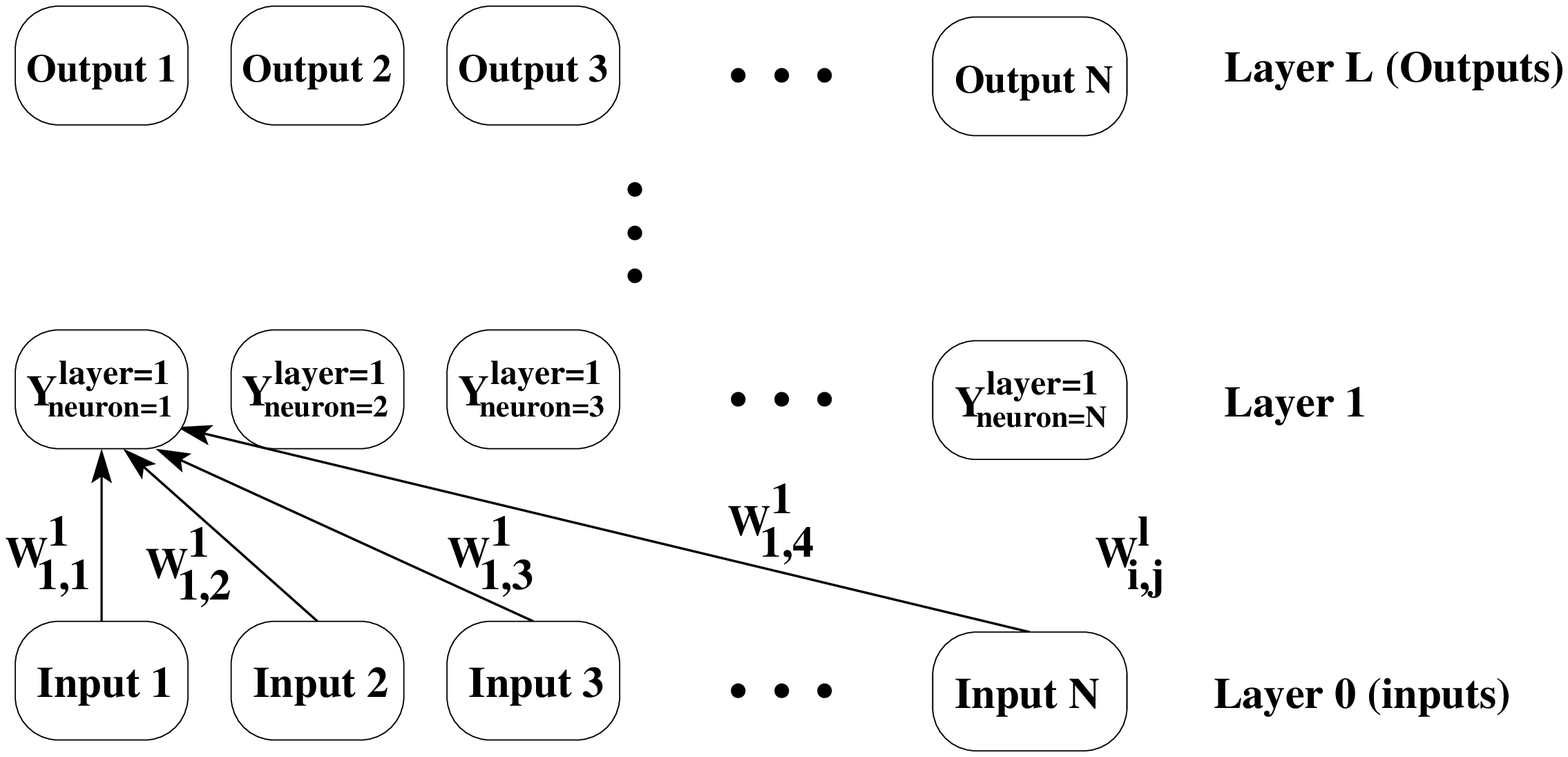}
\caption{
\label{fig:MLP}
Schematic representation of a multi-layer perceptron. Figure reproduced from
\citeN{SN04d}
}
\end{figure*}

As a further simplification, we shall be concerned only with a particular
network configuration: the {\it multilayer perceptron}. In this
configuration the neurons are arranged in successive layers. Each neuron in
any given layer has connections to all neurons in the previous layer. No
connections are allowed between non-successive layers. An arbitrary number of
intermediate layers may exist between the inputs and outputs. These are
usually referred to as ``hidden'' layers in the ANN literature. A schematic
representation of a multi-layer perceptron with hidden layers is given in
Fig~\ref{fig:MLP}. 

When a set of input data is introduced in the ANN, it is propagated forward
according to the following rule:
\begin{equation}
\label{ANNprop}
Y^l_n = f_l \left ( \sum_{j=1}^N W^l_{n,j} Y^{l-1}_j + \beta^l_n \right ) \, .
\end{equation}
In Eq~(\ref{ANNprop}), $Y^l_n$ represents the contents of neuron $n$ in layer
$l$, $W^l_{n,j}$ is the synaptic weight of the connection between this neuron
and neuron $j$ in layer $l-1$ and $\beta^l_n$ is a bias level. We consider
the input neurons as layer $l=0$.
One or more of the ANN layers may have a non-linear ``activation function''
$f_l(x)$. The ANNs used in this work all have:
\begin{eqnarray}
\label{activation}
f_l(x) & = & x  \, \, \, {\rm (for \, linear \, layers)} \, , \nonumber \\
f_l(x) & = & a \tanh (b x) \, \, \, {\rm (for \, non-linear \, layers)} \, .
\end{eqnarray}

The hyperbolic tangent is a common choice for the non-linear activation
function in many ANN applications and has been adopted here for that
reason. Other suitable functions are likely to exist, but will not be pursued
here. After some experimentation with a simplified problem, the parameters
$a$ and $b$ have been set to $1.72$ and $0.67$, respectively. 

Conceptually, a multilayer perceptron may be viewed as a non-linear
mapping ${\bf F}$ between two multi-dimensional spaces. We can write down:
\begin{equation}
{\bf o} = {\bf F} ({\bf x}) \, ,
\end{equation}
where ${\bf x}$ is an N-dimensional input vector and ${\bf o}$ is an
M-dimensional output vector. It can be shown (e.g., \citeNP{J90};
\citeNP{BL91}) that a multilayer perceptron with at least one hidden
non-linear layer is able to approximate any continuous multidimensional
non-linear function to any arbitrary precision, provided only that enough
neurons are employed. This is a very interesting mathematical property that
has received considerable attention in the ANN literature, since it 
provides a solid foundation for many applications.

An ANN is fully characterized by its structure and the individual properties
of its neurons. The network structure is defined by a set of parameters such
as the number of layers $N_L$, the number of neurons in a given layer
$N_N(l)$ (where $l$ denotes the layer), and the activation function of each
layer $f_l(x)$ which can only be one of the two options given explicitly in
Eq~(\ref{activation}) above. According to the choice of $f_l$ 
we shall use the terminology of linear/non-linear layers.

The other set of parameters that completely defines an ANN consists of the
neural biases and synaptic weights ($\beta^l_n$ and $W^l_{n,j}$ in
Eq~[\ref{ANNprop}]) for each single neuron. 

The most important difference between these two sets of parameters is that
the network structure is established {\it a priori} for a given problem and
then kept fixed throughout. The neuron properties (synaptic weights and
biases), however, are adaptive parameters that are continuously adjusted
during the ANN training process (described below) to optimize the network
performance. Although there are algorithms that modify the network structure
during the training process seeking the optimum configuration, I have not
made use of such techniques in the present work. The various ANN structures
used here are the result of a number of trial-and-error experiments. A
detailed study of the impact that the ANN structure has on its performance
would be tremendously expensive from a computational point of view due to the
long processing times involved in training a single network (typically
several days running on a modern workstation).

The ANNs used for the calculations in this work have two linear and two
non-linear layers interleaved. All layers, including the inputs, contain the
same number of neurons (rectangular ANNs). The exception to this is the AANN
of \S\ref{sec:preproc} which has a smaller number of neurons in one of the
hidden layers. 

\subsection{Training a network}
\label{sec:training}

The previous section described the structure of a typical ANN with
particular emphasis on the case of the multilayer perceptron, because this is
the type of ANN employed here. However, nothing has been said thus far about
the particular 
values that should be given to $W^l_{n,j}$ and $\beta^l_n$. It is obvious
that the behavior of the ANN will be dictated by these parameters, which
define the actual mapping ${\bf F}$ performed by the ANN. 

The optimal synaptic parameters are obtained by means of a ``training''
process. For 
the training we need a large set of input vectors $x^t_i$ and the solutions
or outputs $o^t_j$ that we wish the ANN to find (for this reason the outputs
are sometimes referred to as ``targets''). To fix ideas, suppose that we are
training an ANN to take a set of observed spectral profiles as inputs and
return as outputs some parameters of the atmosphere in which the profiles
originated (e.g., temperature, velocity or magnetic field). In this case, the
$x^t_i$ are the observable quantities and the $o^t_j$ are the atmospheric
conditions (in practical situations, one should first pre-process the
inputs and outputs as explained below; therefore these parameters are related
to the actual physical quantities but they are not the quantities
themselves). 

It might seem that, in order to generate a suitable training set one needs
to invert the $x^t_i$ in order to produce the $o^t_j$, which would require
the use of some sort of inversion procedure. However, this is not
necessarily the case. One could start with the atmospheric parameters
(from which the $o^t_j$ are calculated) and then solve the forward problem to
synthesize spectral profiles (thus obtaining the $x^t_i$). 

When the input data are forward-propagated through the ANN we obtain a set of
outputs $o_j$ which are, in general, different from the targets $o^t_j$. The
training algorithm seeks to minimize the difference between the outputs $o_j$
and the targets $o^t_j$ over the entire training set, e.g. in a least-squares
sense by minimizing $\chi^2=\sum_j (o_j - o^t_j)^2$. The most widely used
procedure to accomplish this task is the {\it back-propagation} algorithm,
which performs a non-linear least-squares minimization of the distance
between the network output and the targets (\citeNP{RHW86}).

The back-propagation algorithm needs a starting guess for the synaptic
parameters. I have found that a random initialization with a normal
distribution works well for the applications described here. The amplitude of
the distribution is set to 1/$N_N(l)$. This is done to ensure that the
$Y_n^l$ are of the order of 1. Otherwise one risks entering the saturation
regime of the non-linear activation function. For the same reason it is
important to pre-process the inputs and outputs with a linear transformation
to bring them within the [-1,1] interval (or at least within that order of
magnitude). This has been done here by subtracting the mean value and
normalizing to the standard deviation of each parameter over the entire
dataset. 

For a sake of simplicity, the ANNs used in this work have been trained to
retrieve the magnetic field strength only. It is straightforward to train
similar networks to retrieve any other model parameters that are deemed
relevant. An alternative approach would be to train one single ANN to
retrieve all the parameters. Unfortunately, numerical experimentation
suggests that a much larger number of neurons is necessary if one wishes to
achieve a comparable accuracy in each individual parameter, which increases
the computational cost very rapidly. Therefore, it is probably more
efficient to train a separate ANN for each parameter. In this manner the
computing time for the training and the subsequent inversions scales linearly
with the number of parameters. The additional expense for inverting more than
one parameter is negligible, considering that a typical inversion such as the
ones presented below take only a matter of seconds (compared to several hours
for a full Milne-Eddington inversion).

\subsection{The training and validation sets}
\label{sec:trainset}

The ANNs are trained in successive epochs. A batch of 15000 profiles are
synthesized at each epoch of the training and presented to the network. The
use of many batches helps to minimize ``overfitting'', which is a common
problem  encountered in these applications. Overfitting makes an ANN
reproduce noise 
or other irrelevant features of the dataset and lose generalization
ability. In fact, accuracy and generalization ability are often opposing
atributes and one needs to find an adequate compromise between the two. This
is discussed in more detail in the following sections.

The Milne-Eddington approximation is used for all the spectral synthesis
calculations in this 
paper. This approximation considers that all the relevant parameters that
enter the radiative transfer equation (magnetic field, line strength, Doppler
width, and damping) are constant with height in the line formation
region, except for the source function which varies linearly with optical
depth (see, e.g., \citeNP{LdI92} for details). With these assumptions the
solution to the radiative transfer equation is analytical, which alleviates
considerably the burden of computing thousands of training spectral line 
profiles. In addition to these atmospheric parameters I consider a ``filling
factor'' ($\alpha$) of the magnetic element, which is used to treat spatially
unresolved fields. The magnetic profile is multiplied by $\alpha$ and added a
quiet-Sun profile multiplied by $(1-\alpha)$.

\begin{figure*}
\plotone{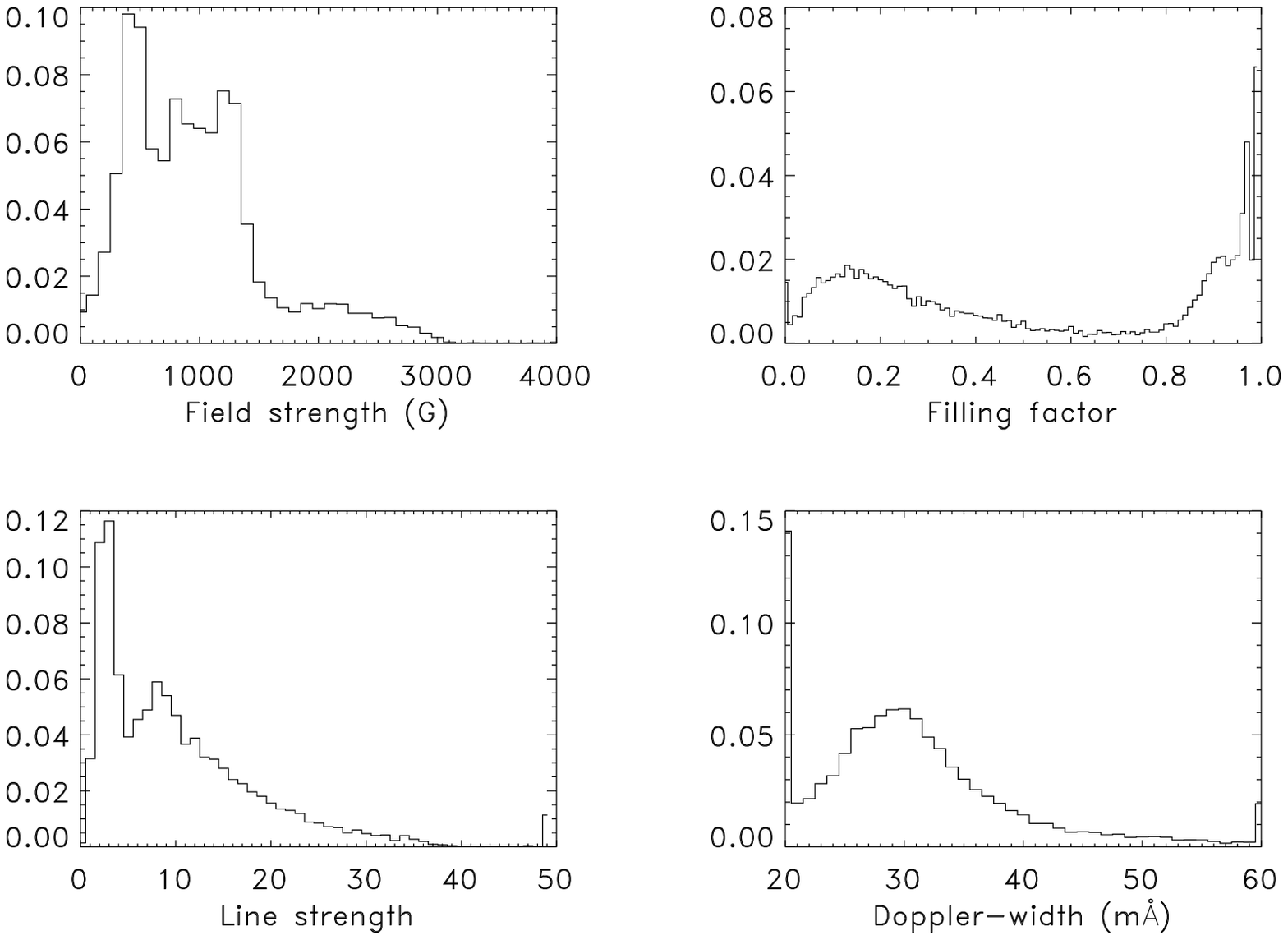}
\caption{
\label{fig:histograms}
Histograms of the distribution of some atmospheric parameters used for the
synthesis of training data.
}
\end{figure*}

\begin{figure*}
\plotone{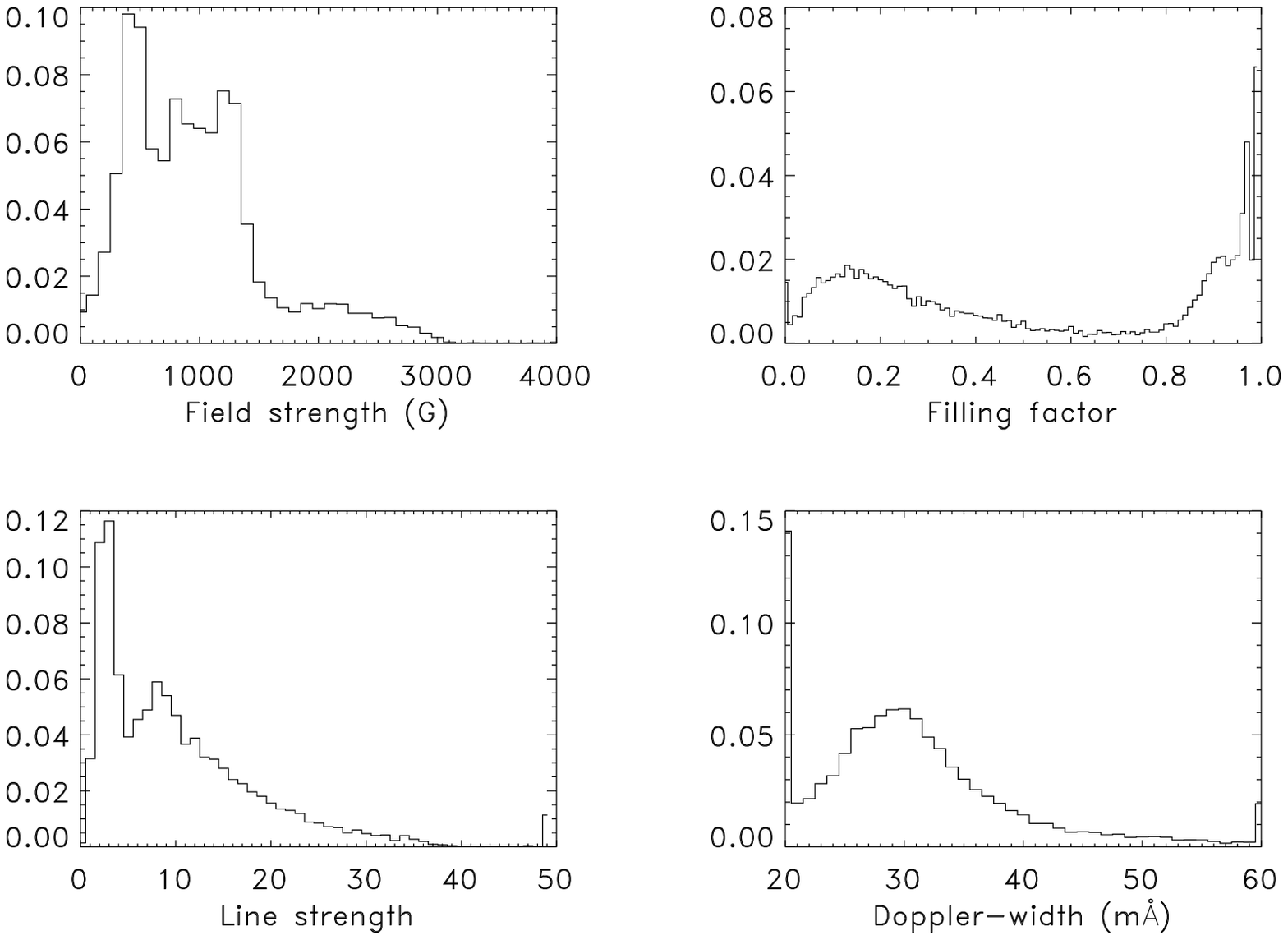}
\caption{
\label{fig:correl}
Correlations between the magnetic field strength and some other atmospheric
parameters used for the synthesis of training data.
}
\end{figure*}

The starting model atmospheres for the spectral syntheses are random but
based on a distribution obtained from actual solar data. I have used existing
observations in the archives of the Advanced Stokes Polarimeter (ASP,
\citeNP{ELT+92}; \citeNP{L96}) to obtain a realistic distribution of the
Milne-Eddington parameters. Histograms of some of the most relevant
parameters are shown in Fig~\ref{fig:histograms}. The various atmospheric
parameters are not independent of one another. Fig~\ref{fig:correl} shows the
correlations existing between some relevant parameters. 

The use of solar distributions has the advantage of optimizing the 
network performance for the observations that one expects to find in the real 
Sun (at least in a statistical sense). 

The profiles calculated are those of the \ion{Fe}{1} pair of lines near
6302~\AA . The spectra are sampled in 52~m\AA \, bins, about a
factor of 4 lower than typical ASP observations. The wavelength range where
two telluric lines are present in actual observations is removed from the
synthetic profiles. 

\begin{figure*}
\plotone{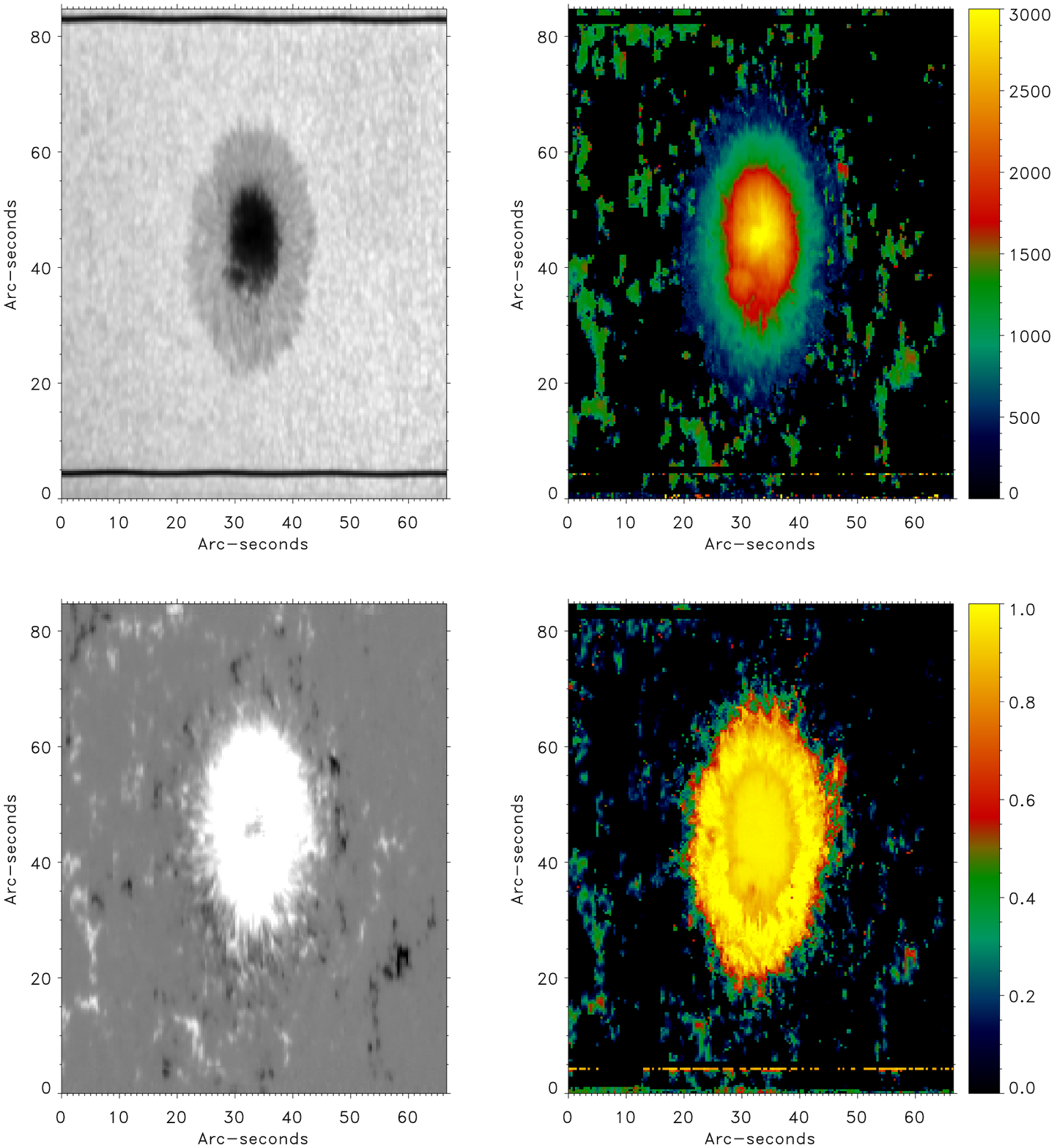}
\caption{
\label{fig:maps}
Observed region. Upper left: Continuum intensity. Upper right: Intrinsic
field strength. Lower left: Degree of circular polarization (saturated at
$\pm$10\%). Lower right: Filling factor of the magnetic element in the pixel
($\alpha$). 
}
\end{figure*}

Once the ANN has been trained, two different tests are used to assess its
performance. The first test uses synthetic profiles obtained from
a Milne-Eddington inversion of ASP observations. The inversions were carried
out by \citeNP{LTB+98} using the code developed by \citeN{SL87}. 
The observed region can be considered rather
typical and contains significant areas of quiet Sun, a fairly round sunspot
and some plage, and has a fairly good spatial resolution ($\simeq$1'') 
consistently during the entire scan. Fig~\ref{fig:maps} shows several maps of
the observed region.

The test with synthetic profiles has the
advantage that the sought solution is known beforehand. Both the training
and the validation data have been produced using the Milne-Eddington
approximation, so the models are consistent physically. Therefore, the errors
obtained can be ascribed directly to the ANN performance.

Unfortunately, the performance of an ANN may be very different when it is
applied to simplified synthetic profiles or to real observations. It is then
crucial to consider a second test in which the validation data are
observations. For this purpose I have used the spectral profiles from the ASP
dataset described above. The ANN outputs are compared to the models resulting
from the Milne-Eddington inversion to estimate the errors. 

\section{The direct approach}
\label{sec:direct}

Let us first consider the most straightforward approach, which is to train
the ANN with synthetic profiles as explained in \S\ref{sec:trainset}. Each
input vector has a total of 80 elements, 
corresponding to 4 Stokes parameters, 2 spectral lines and 10 spectral
samples per line. Output vectors have only 1 element, namely the intrinsic
magnetic field strength (see last paragraph of \S\ref{sec:training}).

Some basic pre-processing is applied to the synthetic profiles before they
are presented to the ANN. This is intended to reduce the dimensionality of
the problem by removing trivial transformations. The following procedures are
applied:

\begin{itemize}
\item The global bulk velocity is removed by shifting all profiles so that
  their Stokes~$I$ ``center of symmetry'' are at the same position. The
  center of symmetry is defined so that the sum of quadratic differences of
  symmetric points in the line profile are minimal. 
\item Random noise is added to all profiles. The noise has a normal
  distribution and an amplitude of 10$^{-3}$ times the continuum intensity.
\item All profiles are normalized to their respective continuum intensity. 
\item The spectral ranges of the two telluric lines are removed.
\item The mean and standard deviation of each spectral sample over the entire
  training set are computed. These are then used to normalize the inputs so
  that they are of the order of 1. The same is done for the output magnetic
  fields. 
\end{itemize}

The ANN is trained in successive epochs with batches of 15000 profiles. For
each epoch, the back-propagation algorithm is applied until 500 iterations
are performed or the last 50 iterations do not result in further
improvement. After a training epoch, the ANN is presented with 500 new
profiles that were not included in the training set. This validation set is
used to estimate the performance of the network when presented with new
data. The training process is repeated in successive training epochs with
different sets until the validation error converges (i.e., it no
longer decreases with additional training batches).

\begin{figure*}
\plotone{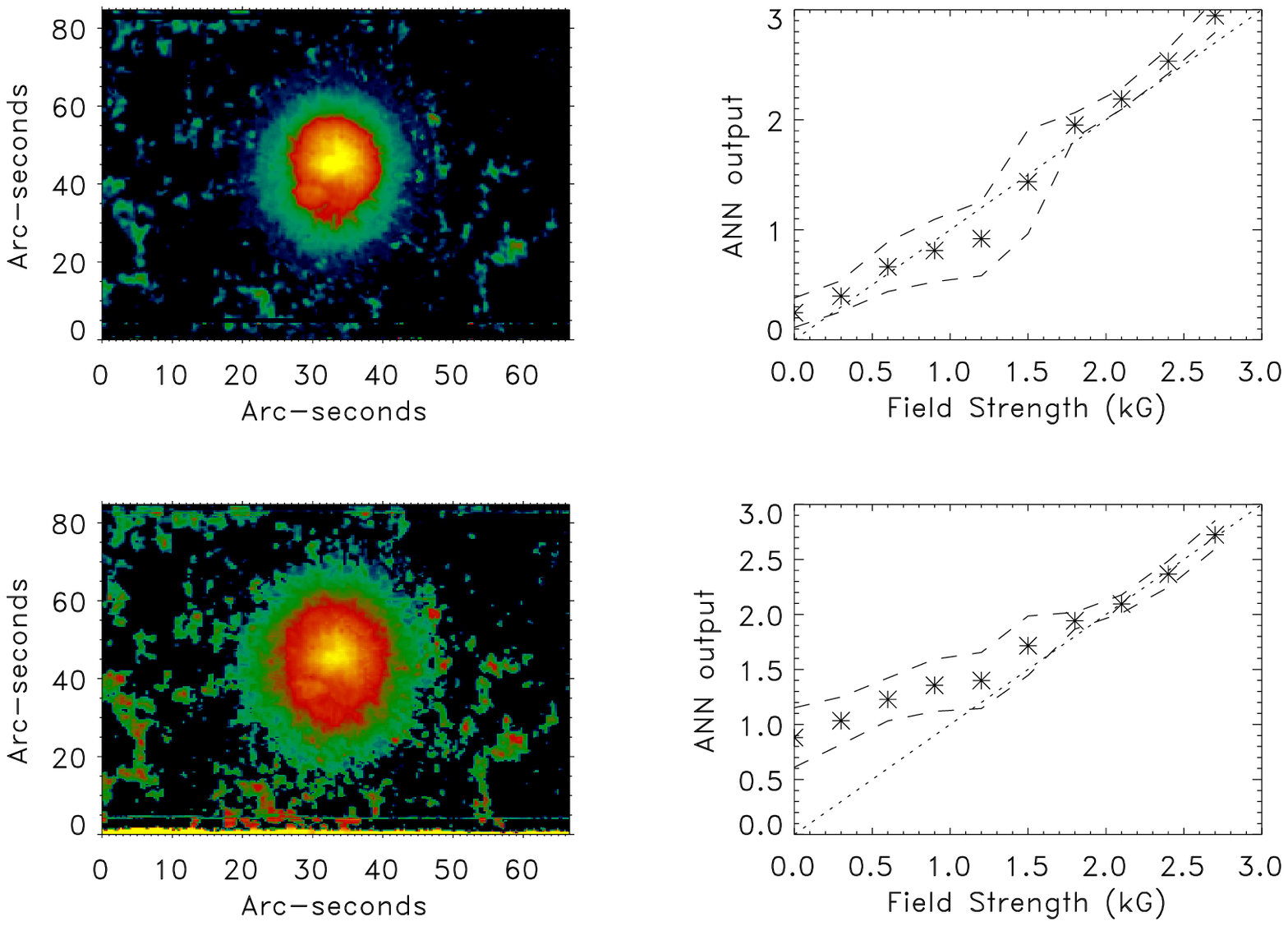}
\caption{Results from ANN inversions with the direct approach. Top:
  Tests with synthetic data. Bottom: Tests with real observations. Left: Maps
  of the magnetic field stregnth retrieved by the ANN. Right: Comparison
  plots between the ANN outputs and the magnetic field from the
  Milne-Eddington inversion. Asterisks: mean value in each bin. Dashed line:
  Standard deviation of the scatter in each bin. Dotted line: Diagonal of the
  plot. 
\label{fig:simplistic}
}
\end{figure*}

After the time-consuming training process ($\sim$2 days), the ANN was tested
with both synthetic and observed ASP data. The results are presented in
Fig~\ref{fig:simplistic}. The performance with synthetic data is very
good, as seen in that figure. However, when faced with actual observations,
this ANN is unable to provide accurate field estimates below $\sim$1~kG. 

The problem with observed profiles is that they exhibit conspicuous features
that are not present in the training set. One has moderate to strong
asymmetries and other departures from the ideal Milne-Eddington profiles
that are unknown to the ANN. This is not so serious in the case
of traditional least-squares inversions because those codes find the solution
that is closest (in a least-squares sense) to the observation. However, the
situation is very different with ANNs. Basically, our network is a
multi-dimensional interpolating function, with the training set representing
the gridpoints that the interpolating function mimics. When we introduce a
point in the ANN that is outside the domain of the training data, the network
is forced to perform an extrapolation instead of an interpolation.

There are two different approaches that one might take to overcome this
issue. The first one is to take the observed profiles and pre-process them in
order to
bring them onto the Milne-Eddington hypersurface (see \citeNP{SN03} for
details). By doing this one is 
effectively looking for the closest Milne-Eddington profile to the
observation. This Milne-Eddington profile is then the one that will be
inverted by the ANN. The second approach is to ``regularize'' the network to
make it tolerant to small deviations of the profiles from the ideal
shape. Both of these strategies are explored in the next sections.

\section{Pre-processing with auto-associative networks}
\label{sec:preproc}

Let us consider first the approach of ``projecting'' the observed profile
vector onto the hypersurface defined by synthetic Milne-Eddington
profiles. This is implicitly done by traditional least-squares fitting codes,
which find the Milne-Eddington profile that is closest to the observation. 

\begin{figure*}
\plotone{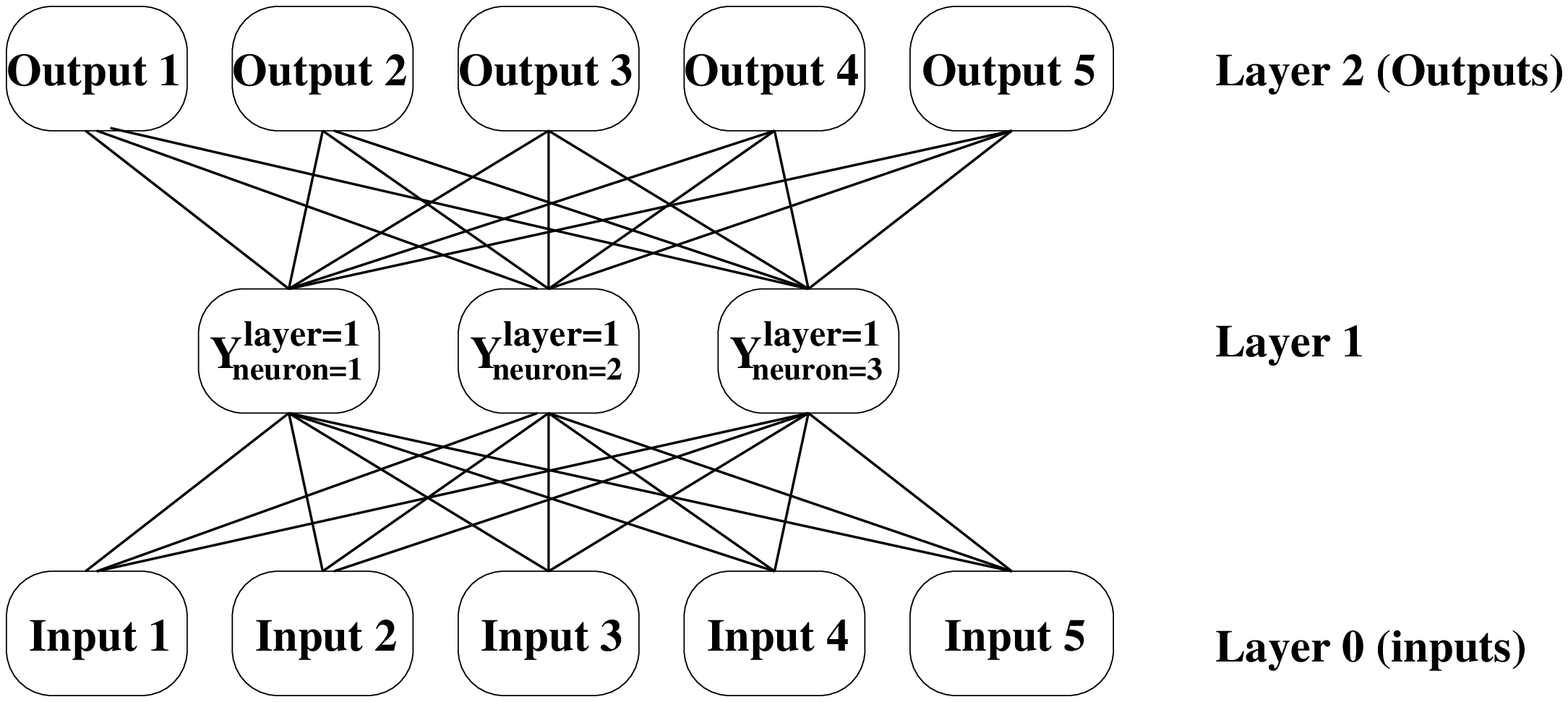}
\caption{Schematic representation of an auto-associative neural
  netowrk. Figure reproduced from \citeN{SN04d}
\label{fig:AANN}
}
\end{figure*}

In the context of ANN inversion, the vector projection is not implicit in the
method, and this needs to be done explicitly (e.g. as a pre-processing of the
input data). In this section I explore a procedure by which an {\it
auto-associative} neural network (AANN) is used for pre-processing. A
previous paper (\citeNP{SN04d}) demonstrated the use of AANNs to decompose
and reconstruct spectral Stokes profiles. The reader is referred to that
paper for details but a brief explanation of AANNs is given here for
completeness. An AANN is simply a neural network that is trained with targets
equal to the input data ($o^t_i = x^t_i$). (Obviously this requires the same
number of neurons in the input and output layers.) An AANN has at least one
hidden layer (the ``bottleneck'' layer) that has fewer neurons than the
input/output layers (see Fig~\ref{fig:AANN}). 

For the application presented here we shall depart slightly from the
conventional 
definition of AANNs, but the underlying principle remains the same. One
starts by synthesizing a large number of Milne-Eddington profiles that will
be used as targets for the AANN. These profiles are then distorted by
artificially adding asymmetries, molecular lines with random positions and
amplitudes (like those found in the umbra of sunspots) and noise. The
quiet Sun intensity profile used for the spatially-unresolved non-magnetic
surrounding is randomized so that it is slightly different for each training
datapoint. The relative velocity between magnetic and non-magnetic elements
is random with an amplitude of 2 km~s$^{-1}$. Finally, both the perturbed and
original profiles undergo basic pre-processing and normalization as explained
in \S\ref{sec:direct}. 

The AANN is trained using the perturbed profiles as inputs and the original
Milne-Eddington ones as targets. The bottleneck layer of this network has only
11 neurons, which is also the number of free parameters in the atmospheric
model. In principle it should be possible for the AANN to extract a set of 11
parameters from the profiles from which these can then be reconstructed.
In practice, however, there may be a significant loss of information
due to the limited number of neurons present between the inputs and the
bottleneck layer, which in turn limits the complexity of the non-linear
transformations that the AANN can do. 

Moreover, the problem we are dealing with here is complicated because
the inputs have been distorted. This means that the AANN needs to find a
suitable set of 11 ``features'' in the distorted spectra from which it
can reconstruct a similar Milne-Eddington profile. Effectively, the AANN is
doing the profile projection mentioned earlier, or at least an approximation
to it.

Once the AANN is properly trained, we can construct another network that will
take the profiles pre-processed by the ANN and do the actual inversion with
them. Notice that the ``inverting'' ANN does not really need to take the
full output vector from the AANN. We can simply take the 11-parameter vector
in the bottleneck layer. This are the parameters that the AANN has determined 
contain all the necessary information to reconstruct the Milne-Eddington
profiles. 

The full data inversion strategy proposed in this section is as follows. We
first train an AANN as explained above with distorted profiles as inputs and
Milne-Eddington profiles as outputs. Once this network has been trained we
shall not make any further use of the layers behind the bottleneck layer. We
then construct a second ANN to do the inversion. To generate the training set
for this new ANN we must first pre-process all the training profiles with the 
AANN. Each training profile is propagated through the AANN, but only up to
the bottleneck layer. The features in the bottleneck layer are used as inputs
to the inverting ANN, whereas the targets are the magnetic field strength
properly normalized as in \S\ref{sec:direct}.

\begin{figure*}
\plotone{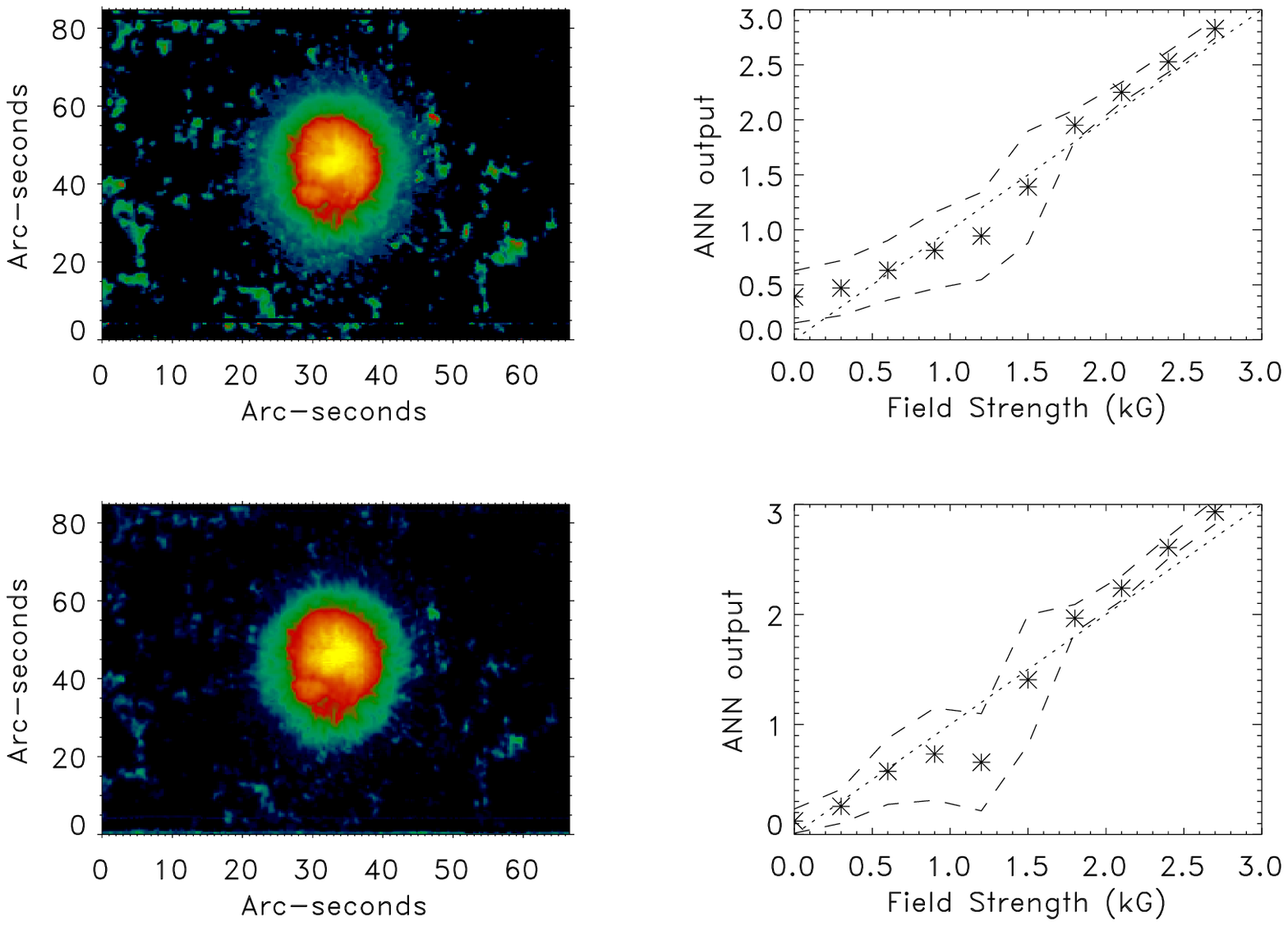}
\caption{Results from ANN inversions with the pre-processing approach. Top:
  Tests with synthetic data. Bottom: Tests with real observations. Left: Maps
  of the magnetic field stregnth retrieved by the ANN. Right: Comparison
  plots between the ANN outputs and the magnetic field from the
  Milne-Eddington inversion. Asterisks: mean value in each bin. Dashed line:
  Standard deviation of the scatter in each bin. Dotted line: Diagonal of the
  plot. 
\label{fig:preproc}
}
\end{figure*}

The results of the tests with this complex combination of inputs
pre-processing and data inversion are shown in Fig~\ref{fig:preproc}. The
performance with synthetic profiles has degraded somewhat with respect to the
case of the direct approach (Section~\ref{sec:direct}). However, we can see
that now the accuracy of inversions with observed and synthetic profiles are
very similar. The errors are reasonably small, close to (and sometimes
smaller than) 100~G except for the range between 1~kG and 1.5~kG. This region
is complicated because it contains most of the quiet Sun fields, which
usually have small filling factors. Inverting these profiles is 
difficult because the polarization signals are typically much weaker (in
spite of the fact that the intrinsic field may be strong) and there is
rarely any linear polarization at all. The ANN-based inversion tends to
underestimate these fields, although the correct solution is still within the 
1-$\sigma$ scatter in the plot.

\section{Network regularization}
\label{sec:regul}

The third strategy proposed in this paper is based on the concept of
``regularization'', which aims at making the ANN tolerant to small departures
from the model. This is usually accomplished by training ANNs with noisy 
data. In our problem, however, the difficulties reside not only in the noise
but more importantly in spurious features such as asymmetries or molecular
lines that effectively produce distortions in the profiles. 

Regularization may reduce the accuracy of an ANN, especially when working with
high-quality data. However, it makes it more robust and improves its
generalization ability. One usually needs to find a suitable compromise
between these two qualities.

This section explores the applicability of a regularized ANN for the
inversion of spectral Stokes data. I have trained a single ANN with input
profiles that have been distorted exactly as those for the AANN in
\S\ref{sec:preproc} and the magnetic fields as targets. The goal is to
combine the two steps (pre-processing and inversion) in one. The overall
training time is somewhat longer than in previous cases (about 50\% longer
than the direct approach).

\begin{figure*}
\plotone{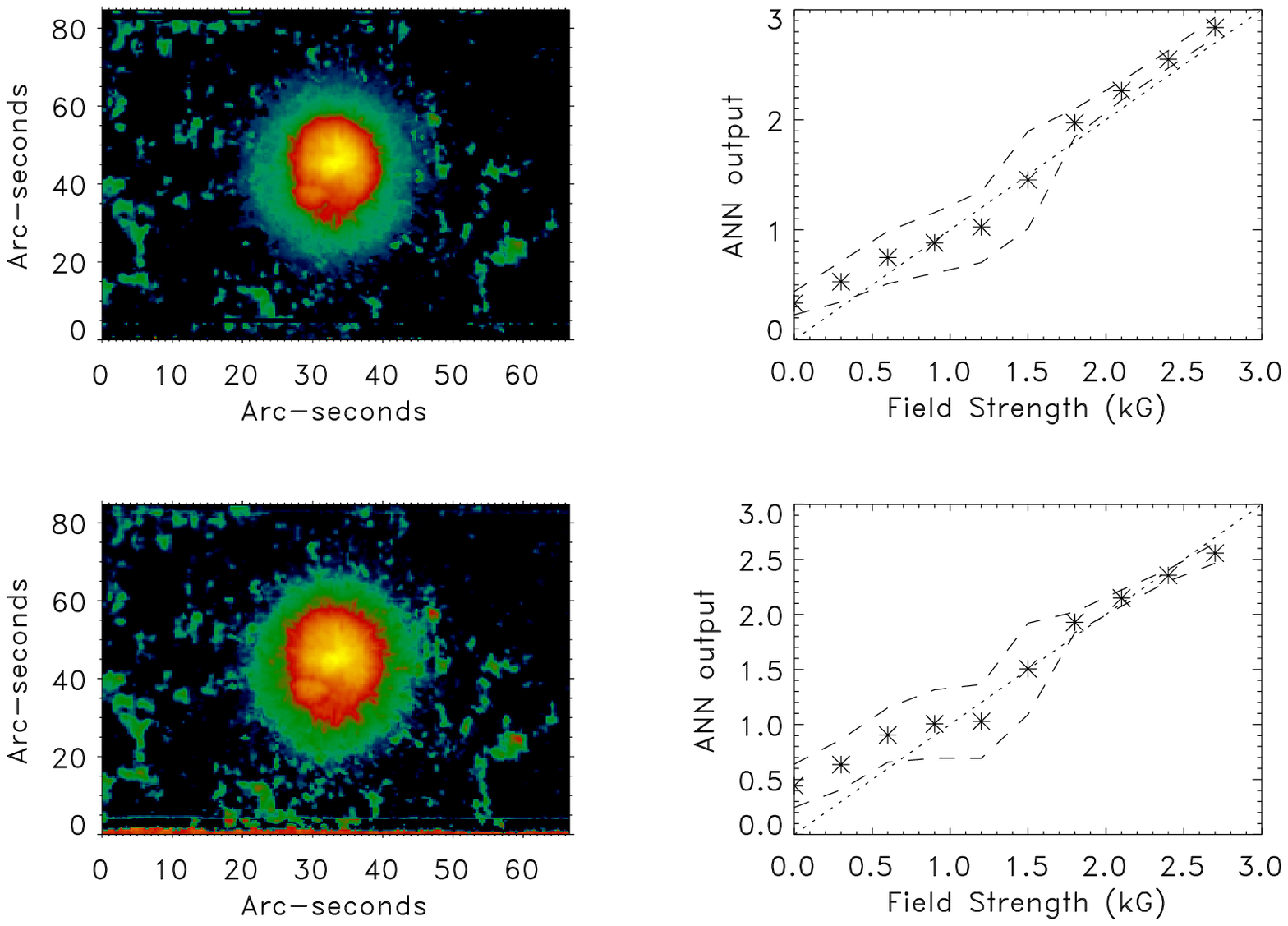}
\caption{Results from ANN inversions with the regularization approach. Top:
  Tests with synthetic data. Bottom: Tests with real observations. Left: Maps
  of the magnetic field stregnth retrieved by the ANN. Right: Comparison
  plots between the ANN outputs and the magnetic field from the
  Milne-Eddington inversion. Asterisks: mean value in each bin. Dashed line:
  Standard deviation of the scatter in each bin. Dotted line: Diagonal of the
  plot. 
\label{fig:regul}
}
\end{figure*}

The results of applying this ANN to the test data are shown in
Fig~\ref{fig:regul}. As before, the network performance with synthetic data
is slightly worse than that of the direct approach. However, we gain an
important benefit in that it can invert observations practically with the
same accuracy as synthetic data. Compared to the case of \S\ref{sec:preproc},
this particular ANN performs only a little worse for the weaker fields (below
$\sim$1~kG) which are slightly overestimated. These points are found mostly
in the outer penumbra (some of them also in the quiet Sun, but they are
few). The quiet Sun points between 1~kG and 1.5~kG are better retrieved than
in the previous case. There is still a slight underestimation, but the effect
is small.

\section{Conclusions}
\label{sec:conc}

This paper shows that ANNs are a viable alternative to least-squares fitting
for the routine analysis of large amounts of data. While their accuracy is
somewhat lower, the ability to process much larger datasets will
probably present advantages for some application. The CPU times required for
the ANN inversions presented here are $\sim$10~seconds, compared to
$\sim$5~hours for the original Milne-Eddington inversion.

Furthermore, one does not have to be concerned with the algorithm finding
secondary minima or not converging. Strictly speaking, the ANN inversion 
is not an inverse problem. It is rather a case of interpolating a
multidimensional function that maps a set of gridpoints from the space of
spectra into the space of models.

It is important to emphasize that pattern recognition techniques are not
meant to replace traditional least-squares fitting algorithms, but to
complement them. For detailed studies of a smaller data subset or individual
profiles, or if one needs to consider more realistic model atmospheres (with
line-of-sight gradients, Non-LTE effects, etc), it is still necessary to use
a least-squares inversion. 

\acknowledgments
The author is grateful to B.W. Lites for providing the observations used for
the tests in this paper and to T. Carroll for many fruitful discussions.


\end{document}